
\magnification=\magstep1
\rightline{NUB-3130}
\vskip 1cm
\centerline{\bf Comment on ``EPR without `collapse of the wave function' ''}
\bigskip
\centerline{Y. N. Srivastava and A. Widom}
\centerline{Physics Department, Northeastern University, Boston MA, USA }
\centerline{Physics Department \& INFN, University of Perugia, Perugia, Italy}
\centerline{and}
\centerline{E. Sassaroli}
\centerline{Center for Theoretical
Physics, Massachusetts Institute of Technology, Cambridge MA }
\centerline{Physics Department \& INFN, University of Perugia, Perugia, Italy}
\vskip 2.0cm

\centerline{\it ABSTRACT}
\bigskip
The implications of the many proper time amplitudes for the process $\Phi
\rightarrow K^o+\bar{K}^o$ was discussed in a recent letter of B.
Kayser and L. Stodolsky who predicted a small correction $(\sim 5\%)$ to the
usually predicted phase oscillation frequency for forthcoming $\Phi $
factory experiments. Based on previous work (also using the many proper
times formalism), we predicted a much larger correction factor $\approx 2$.
This correction factor is important in that it allows for a crucial future $%
DA\Phi NE$ experimental test of the many proper times amplitude formalism.
\vskip 6cm
\noindent
$\dagger $ Work supported by DOE and INFN
\vfill \eject

In a recent letter$[1]$, B. Kayser and L. Stodolsky discussed issues
concerning quantum interference of amplitudes and their conclusions ({\it
bar one}) are remarkably similar to those of our previous work$[2-8]$.

The issues are as follows: (i) The notion of the ``collapse of the wave
function'' (or equivalently the so called ``projection postulate'') is not
generally valid for reasons of Lorentz symmetry, and (ii) Each internal
propagator of a Feynman diagram requires its own proper time
(many proper time amplitudes)$^\dagger$(1).

The above ideas were applied to the $DA\Phi NE$ process
$$
\Phi \rightarrow K^0+\bar{K}^0.\eqno(1)
$$
An example of two interfering common channel (secondary vertex) amplitudes
are shown below; i.e.
\medskip
$$
\matrix{\pi^0 &\ &\ &\ &\ &\ &\ &\ &\pi^+ \cr
\nwarrow &\ &\ &\ &\ &\ &\ &\ & \nearrow \cr
\ &-&K_{S,left}&-&\Phi &--&K_{L,right}&--&\ \cr
\swarrow &\ &\tau_{S,left} &\ &\ &\ &\tau_{L,right}&\ &\searrow \cr
\pi^0 &\ &\ &\ &\ &\ &\ &\ &\pi^-
}, \eqno(2a)
$$
\medskip
\noindent and
\medskip
$$
\matrix{ \pi^0 &\ &\ &\ &\ &\ &\ &\ &\pi^+ \cr
\nwarrow &\ &\ &\ &\ &\ &\ &\ & \nearrow \cr
\ &-&K_{L,left}&-&\Phi &--&K_{S,right}&--&\ \cr
\swarrow &\ &\tau_{L,left} &\ &\ &\ &\tau_{S,right}&\ &\searrow \cr
\pi^0 &\ &\ &\ &\ &\ &\ &\ &\pi^-
}.\eqno(2b)
$$
\medskip  \noindent
Note that {\it four} proper times are required to describe these {\it two}
interfering amplitudes. The resulting total amplitude ($K_L$ and $K_S$ are
virtual) has the form
$$
Amp( \Phi \rightarrow (\pi ^0\pi ^0)_{left}+(\pi ^{+}\pi
^{-})_{right}) =
$$
$$
A( K_L\rightarrow \pi ^{+}\pi ^{-}) A( K_S\rightarrow \pi^0\pi ^0)
\exp ( -i\bar{M}_S\tau _{S,left}-i\bar{M}_L\tau _{L,right})
$$
$$
-A( K_S\rightarrow \pi ^{+}\pi ^{-}) A( K_L\rightarrow \pi^0\pi ^0)
\exp ( -i\bar{M}_S\tau _{S,right}-i\bar{M}_L\tau _{L,left}),
\eqno(3)
$$
where $\bar{M}_{L,S}=(M_{L,S}-(i/2)\Gamma_{L,S})$.
The interference phase between the processes in Eqs.(2) is given by
$$
\theta =M_S\tau _{S,left}+M_L\tau _{L,right}-
M_S\tau _{S,right}-M_L\tau _{L,left}. \eqno(4)
$$
With
$$
M=(1/2)(M_L+M_S),\ \ \ \Delta M=(M_L-M_S),\eqno(5a)
$$
$$
\tau_{right}=(1/2)(\tau _{S,right}+\tau _{L,right}),\ \
\tau_{left}=(1/2)(\tau _{S,left}+\tau _{L,left}), \eqno(5b)
$$
and
$$
\Delta \tau_{right}=(\tau _{L,right}-\tau _{S,right}),\ \
\Delta \tau_{left}=(\tau _{L,left}-\tau _{S,left}), \eqno(5c)
$$
we previously found that
$$
\theta = \Delta M(\tau _{right}-\tau _{left})
+M(\Delta \tau_{right}-\Delta \tau_{left}). \eqno(6)
$$

Kayser and Stodolsky erroneously neglected the second term on
the right hand side of Eq.(6). The error turns out to be crucial.
If one neglects the second term on the right hand side of Eq.(6),
then there is no difference between using four times or using the
two times conventionally employed previous to our work. In fact,
it is the second term on the right hand of Eq.(6) that provides a
forthcoming clear experimental $DA\Phi NE$ test of the many time
amplitude formalism. The kinematics are as follows: (i) In the center
of mass frame (i.e. the $e^++e^-\rightarrow \Phi $ rest frame at
$DA\Phi NE$) we have
${\bf p}_L+{\bf p}_S={\bf 0}$ so that $p=|{\bf p}_L|=|{\bf p}_S|$.
(ii) For a long or short $K$ meson $\tau_{L,S}=(M_{L,S}d/p)$ where
$d$ is the center of mass frame distance that the $K$ meson travels.
(iii) Thus $\tau_{L}-\tau_{S}=([M_{L}-M_{S}]d/p)$, or equivalently
$\Delta \tau\approx (\Delta M/M)\tau $. (iv) Finally
$$
M(\Delta \tau_{right}-\Delta \tau_{left})\approx
\Delta M(\tau_{right}-\tau_{left}),  \eqno(7)
$$
and the two terms on the right hand side of Eq.(6) are approximately
equal. With proper units restored,
$$
\theta \approx  (2c^2\Delta M/\hbar )(\tau _{right}-\tau _{left}).
\eqno(8)
$$
We note in passing that the interference phase in Eq.(8) contains
light speed $c$ (relativity). Any attempt to derive $\theta $ using
purely non-relativistic quantum mechanics will be inadequate.
\vfill \eject
\centerline{\bf FOOTNOTE}
\par \noindent
(1) Compare footnote 9 of[1] with Eq.(48) of[6].
\bigskip
\centerline{\bf REFERENCES}
\bigskip
\par \noindent
[1] B. Kayser and L. Stodolsky, {\it Phys Lett B}{\bf 359}\ (1995) 343.
\par \noindent
[2] A. Widom, E. Sassaroli, and Y. N. Srivastava,``Two-time Formula for
DA$\Phi$NE'', Proceedings of the II EURODA$\Phi$NE Collaboration Meeting,
Edited by L. Maiani, G. Pancheri, and N. Paver (1994).
\par \noindent
[3] Y. N. Srivastava, A. Widom, and E. Sassaroli, ``Past, Future and
Elsewhere in Quantum Mechanics'', Proceedings of a Conference on
Phenomenology of Unification From Present to Future, (March 1994),
Edited by G. Diambrini-Palazzi et al, p.265 (World Scientific, 1995).
\par \noindent
[4] Y. N. Srivastava, E. Sassaroli, and A. Widom, ``Two-distance and Two-
time Formula for $K{\bar K}$ and $B{\bar B}$ Decays'', Abstract presented
at the 25th International Conference on High-Energy Physics, 20-
27 July 1994, Glasgow, Scotland.
\par \noindent
[5]  Y. N. Srivastava, A. Widom, and E. Sassaroli,
{\it Phys Lett B}{\bf 344}\ (1995)\ 436.
\par \noindent
[6]  Y. N. Srivastava, A. Widom, and E. Sassaroli, {\it Zeit Phys C}
{\bf 66}\ (1995)\ 601.
\par \noindent
[7]  Y. N. Srivastava, A. Widom, and E. Sassaroli, ``Real and Virtual
Strange Processes'', Proceedings of the Workshop on Physics and Detectors
for DA$\Phi$NE, Laboratori Nazionali di Frascati, Frascati, April 1995
(hep-ph/9507330).
\par \noindent
[8] E. Sassaroli, Y. N. Srivastava, and A. Widom, ``Charged Lepton
Oscillations'', Northeastern University Preprint, (submitted for
publication)(hep-ph/950961).
\par \noindent

\bye